\documentclass[12pt,preprint]{aastex}

\shorttitle{The distribution of H$^{13}$CN in the CSE around IRC\,+10216}
\shortauthors{Sch\"oier et al.}

\begin{document}


\title{The distribution of H$^{13}$CN in the circumstellar envelope around IRC\,+10216}


\author{Fredrik L. Sch\"oier}
\affil{Stockholm Observatory, AlbaNova University Center, SE-106 91 Stockholm, Sweden \\
         Onsala Space Observatory, SE-439 92 Onsala, Sweden}
\email{schoier@chalmers.se}

\author{David Fong}
\affil{Harvard-Smithsonian Center for Astrophysics,  60 Garden Street, Cambridge, MA 02138, USA}

\author{John H. Bieging}
\affil{Steward Observatory, The University of Arizona, Tucson AZ 85721, USA}

\author{David J. Wilner}
\affil{Harvard-Smithsonian Center for Astrophysics,  60 Garden Street, Cambridge, MA 02138, USA}

\author{Ken Young}
\affil{Harvard-Smithsonian Center for Astrophysics,  60 Garden Street, Cambridge, MA 02138, USA}

\author{Todd R. Hunter}
\affil{National Radio Astronomy Observatory, 520 Edgemont Rd, Charlottesville, VA 22903, USA}




\begin{abstract}
H$^{13}$CN $J$\,=\,8\,$\rightarrow$\,7 sub-millimetre line emission produced in the circumstellar envelope around the extreme carbon star IRC+10216 has been imaged at sub-arcsecond angular resolution using the SMA. 
Supplemented by a detailed excitation analysis the average fractional abundance of H$^{13}$CN in the inner wind ($\lesssim$\,5\,$\times$\,10$^{15}$\,cm) is estimated to be about 4\,$\times$\,10$^{-7}$, translating into a total HCN fractional abundance of 2\,$\times$\,10$^{-5}$ using the isotopic ratio $^{12}$C/$^{13}$C\,$=$\,50. Multi-transitional single-dish observations further requires the H$^{13}$CN fractional abundance to remain more or less constant in the envelope out to a radius of $\approx$\,4\,$\times$\,10$^{16}$\,cm, where the HCN molecules are effectively destroyed, most probably, by photodissociation. The large amount of HCN present in the inner wind provides effective line cooling that can dominate over that generated from CO line emission. It is also shown that great care needs to be taken in the radiative transfer modelling where non-local, and non-LTE, effects are important and where the radiation field from thermal dust grains plays a major role in exciting the HCN molecules. 
The amount of HCN present in the circumstellar envelope around IRC+10216 is consistent with predicted photospheric values based on equilibrium chemical models and  indicates that any non-equilibrium chemistry occurring in the extended pulsating atmosphere has no drastic net effect on the fractional abundance of HCN molecules that enters the outer envelope. It further suggests that few HCN molecules are incorporated into dust grains. 
\end{abstract}


\keywords{stars: abundances -- stars: AGB and post-AGB -- stars: carbon -- stars: circumstellar matter -- stars: individual (IRC\,+10216) -- stars: mass loss}



\section{Introduction}
IRC+10216 (CW Leo) has for a long time been used as an archetype for carbon stars. The reason for this is the combination of its relative proximity (120\, pc) and high mass-loss rate (on average 1.5$\times$10$^{-5}$\,M$_{\odot}$\,yr$^{-1}$; Sch\"oier \& Olofsson 2001) making it possible to detect, with relative ease, a wealth of different molecular species and allowing for detailed studies of their spatial distributions \citep[e.g.,][]{Keady93, Bieging93, Dayal93, Dayal95, Lucas95, Fong03, Ford04, Young04, Schoeier06c, Fong06}. Most current chemical models of carbon stars are also aimed at explaining various molecular abundances derived for this source from observations \citep[e.g.,][]{Millar94,Willacy98, Doty98, Millar00, Millar01, Agundez06}. However, as demonstrated by \citet{Woods03} in a study of a small sample of high mass-loss rate carbon stars, a large spread in molecular fractional abundances, in some cases more than an order of magnitude, were found among the sources. 

Surprisingly large amounts of oxygen bearing molecules such as, e.g., H$_2$O, H$_2$CO, C$_3$O and SiO have been found in the carbon-rich envelope of IRC+10216 \citep{Melnick01, Hasegawa06, Agundez06, Tenenbaum06, Schoeier06c}. SiO fractional abundance estimates for a large sample of carbon stars also indicates that such `anomalous' chemistries might be a common phenomenon in carbon-rich circumstellar envelopes (CSEs) \citep{Schoeier06a}. Various suggestions as to explain the observed molecular abundances have been made including non-equilibrium chemical processes, Fischer-Tropsch catalytic processes, and evaporation of cometary bodies \citep{Bieging00, Melnick01, Willacy04, Cherchneff06, Agundez06}.

\noindent
One of the molecules that shows least variation in its fractional abundance between carbon stars is HCN, less than a factor of two in the sample of \citet{Woods03}. Furthermore, \citet{Lindqvist00}, imaged a handful of carbon stars, with a large spread in mass-loss rates, in the HCN $J$\,=\,1\,$\rightarrow$\,0 transition using the Plateau de Bure interferometer. Combined with a detailed radiative transfer modelling they found that the HCN abundance in their sample varied by less than a factor of three in normal carbon stars. A possible explanation for this is that the HCN abundance is determined close to the photosphere in near equilibrium chemistry and that the shocked non-equilibrium chemistry occurring in the extended pulsating atmosphere only marginally affects the HCN abundance in a carbon-rich environment. 

Nevertheless, HCN is of great importance for the circumstellar chemistry further out in the envelope where its photodissociation is the start of building more complex carbon chain molecules \citep[e.g.,][]{Cherchneff93a, Cherchneff93b}. In addition, line emission from HCN  has been suggested to be, together with CO, the dominant molecular coolant in carbon rich circumstellar envelopes \citep{Cernicharo96}.

In this paper we present new interferometric observations using the Submillimeter Array\footnote{The Submillimeter Array is a joint project between the Smithsonian
Astrophysical Observatory and the Academia Sinica Institute of Astronomy and
Astrophysics, and is funded by the Smithsonian Institution and the Academia
Sinica.} (SMA) which, together with multi-transitional single-dish observations, help constrain the H$^{13}$CN abundance distribution in the circumstellar envelope around IRC+10216. The observations are described in Sec.~2 and the brightness distribution is discussed in Sect.~3. The excitation analysis of the observations is presented in Sect.~4 and the results are discussed in Sect.~5. The paper is then concluded in Sect.~6.

   \begin{figure*}
    \label{sma_fig}
   \centering{   
   \includegraphics[width=16cm]{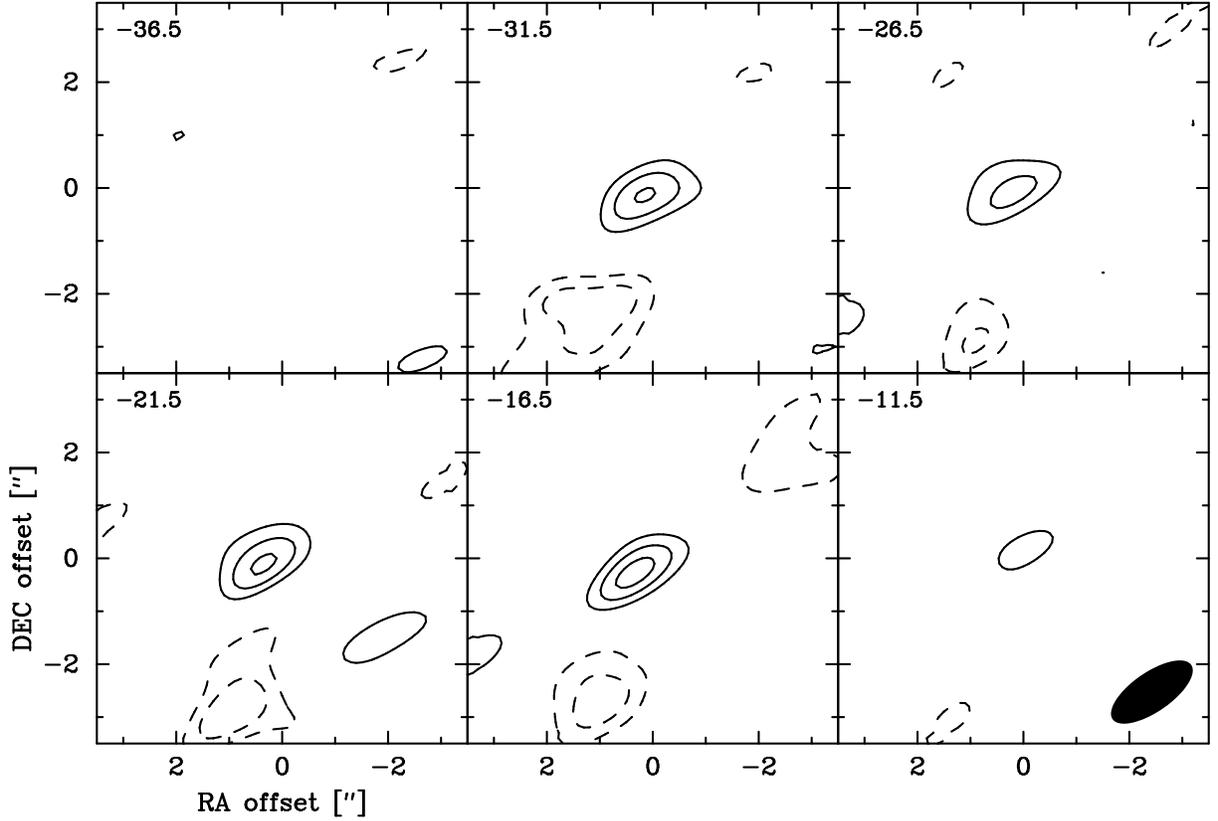}
   \caption{Velocity channel maps of H$^{13}$CN $J$\,$=$\,8\,$\rightarrow$\,7 line emission from IRC\,+10216 obtained using the SMA.  
The contour levels are $15$ Jy~beam$^{-1}$, where $n=-3,-2,2,3,4$ (negative values have dashed contours), and the beam size
is $1\farcs6\times0\farcs7$ with a position angle of $-58\degr$ as indicated in the lower right corner of the bottom right panel.  The velocity channels (given in the
LSR frame and indicated in the upper left corner) have been binned to 5\,km\,s$^{-1}$. The systemic velocity is $-$26\,km\,s$^{-1}$ (LSR) as determined from CO millimeter line observations.
Offsets in position are relative to $\alpha_{2000} = 09^{\rm h} 47^{\rm m} 57\fs39$, $\delta_{2000} = 13\degr 16\arcmin 43\farcs9$.}}
   \end{figure*}

\section{Observations}
\label{sect_obs}
\subsection{The SMA observations}
\label{obs_sma}
SMA observations of the H$^{13}$CN $J$\,$=$\,8\,$\rightarrow$\,7 line at 690.55207\,GHz
were made on 2004 April 1 using the compact array configuration with eight 
6\,m antennas, of which six antennas were equipped with receivers
for 690\,GHz band observations.  Details of the SMA are described by \citet{Ho04}.  
The adopted source position for IRC+10216 was $\alpha_{2000} = 09^{\rm h} 47^{\rm m} 57\fs39$, $\delta_{2000} = 13\degr 16\arcmin 43\farcs9$.
The
weather was good, with a 225\,GHz atmospheric opacity of 0.05 (measured
at the nearby Caltech Submillimeter Observatory) and stable through the night of observations. 
The corresponding opacity at 690.55\,GHz is  0.98 with a zenith transmission of 38\%.
System temperatures at 690\,GHz were in the range from 2000\,$-$\,8000\,K (DSB).  The projected baselines ranged from  17\,$-$\,68\,m (39\,$-$\,157\,k$\lambda$) resulting in a synthesized beam 
of $1\farcs6\times0\farcs7$, with a PA of $-58\degr$ using natural weighting. 
In an aperture synthesis image the beam size is determined by the diameter of the `synthesised aperture', which is twice the maximum baseline in the case of an uniformly sampled $uv$-plane. In what follows we will adopt the term `equivalent distance' to describe this transformation from baseline separation to angular resolution. 
Thus, we expect to obtain usable information on scales as low as $0\farcs$7, corresponding to the longest baselines.

The digital
correlator has a bandwidth of 2\,GHz and the spectral resolution was 0.406\,MHz,
corresponding to a velocity resolution of 0.18\,km\,s$^{-1}$.  Phase and amplitude
calibration were performed with the Jovian moons Callisto and Ganymede.
Bandpass calibration used Callisto and Ganymede as well as Saturn and Uranus.  Observations of
Callisto (assuming 120\,K) provided the flux calibration; the uncertainty of the flux scale is
estimated to be $\approx$\,$\pm$\,30\%.  The data were calibrated using the MIR software
package developed originally for the Owens Valley Radio Observatory and adapted
for the SMA.  The calibrated visibility data were imaged and CLEANed \citep{Hoegbom74} using MIRIAD.
Velocity channel maps of the H$^{13}$CN $J$\,$=$\,8\,$\rightarrow$\,7  line emission obtained by the SMA are presented in Fig.~\ref{sma_fig}.

The actual analysis and comparison with models are carried out
in the $uv$-plane to maximize the sensitivity and resolution of the
data.  

\begin{table}
\caption{H$^{13}$CN single-dish observations of IRC+10216.}
\label{intensities}
$$
\begin{array}{cccccccc}
\hline
\noalign{\smallskip}
\multicolumn{1}{c}{{\mathrm{Transition}}} & 
\multicolumn{1}{c}{{\mathrm{Frequency}}} &  
\multicolumn{1}{c}{E_u} &  
\multicolumn{1}{c}{{\mathrm{Telescope}}}  &
\multicolumn{1}{c}{\theta_{\mathrm{mb}}}  & 
\multicolumn{1}{c}{\eta_{\mathrm{mb}}}  &
\multicolumn{1}{c}{\int T_{\mathrm{mb}}dv}  \\ 
& 
\multicolumn{1}{c}{{\mathrm{[GHz]}}} &
\multicolumn{1}{c}{{\mathrm{[K]}}} & & 
\multicolumn{1}{c}{[\arcsec]} & &
\multicolumn{1}{c}{[\mathrm{K\,km\,s}^{-1}]} \\
\noalign{\smallskip}
\hline
\noalign{\smallskip}
J=1\rightarrow0 &  \phantom{0}86.340  & \phantom{00}4 &  \mathrm{OSO}    & 44 & 0.60 &  121.9 \\
J=3\rightarrow2 &                      259.012  & \phantom{0}25 &  \mathrm{JCMT}  & 19 & 0.69 & 419.8 \\
J=4\rightarrow3 &                      345.340  & \phantom{0}41 &  \mathrm{JCMT}  & 14 & 0.63 & 503.2 \\
J=8\rightarrow7 &                      690.552  &                     149 &  \mathrm{CSO}    & 10 & 0.41 & 405.0 \\

\noalign{\smallskip}
\hline
\end{array}
$$
\end{table}
   \begin{figure*}
      \centering{   
   \includegraphics[width=16.5cm]{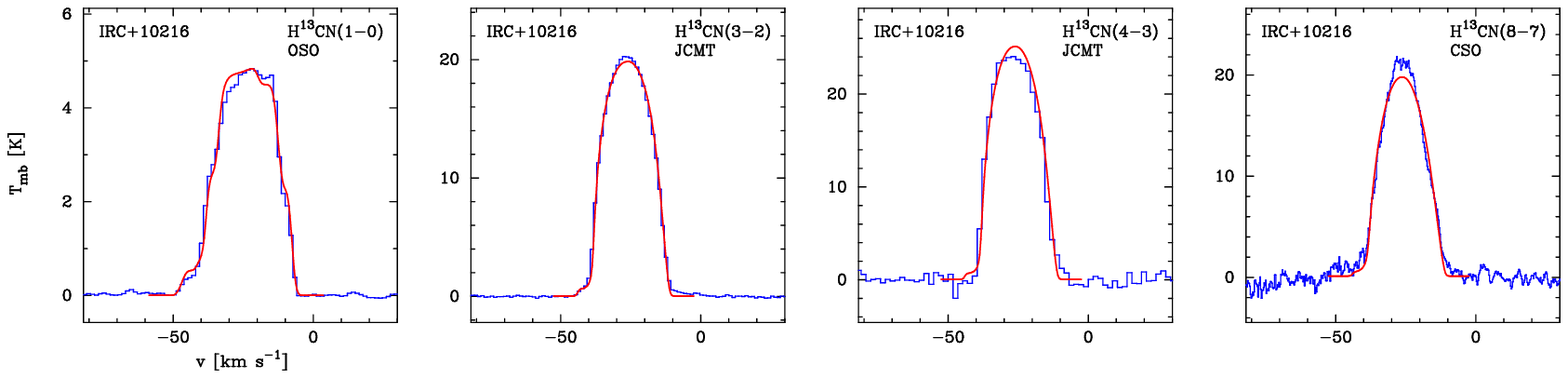}
   \caption{Multi-transition spectra (histograms) of H$^{13}$CN line emission toward IRC+10216. Spectra from the best-fit single-dish model (Model 1; solid lines) using a fractional H$^{13}$CO abundance of 6$\times$10$^{-7}$ and a envelope size of 3$\times$10$^{16}$\,cm are also shown. The hyperfine splitting of the $J$\,=\,1\,$\rightarrow$\,0 transition significantly broadens the line. This effect is explicitly taken into account in the modelling.}
   \label{spectra}}
      \end{figure*}

\subsection{Single-dish observations}
Multi-transition H$^{13}$CN observations of circumstellar emission around IRC+10216 have been performed using several telescopes and are summarized in Table~\ref{intensities}. The individual spectra are shown in Fig.~\ref{spectra}. Observations of $J$\,$=$\,1\,$\rightarrow$\,0 line emission were carried out using the Onsala 20\,m telescope\footnote{The Onsala 20\,m telescope is operated by the  Swedish National Facility for Radio Astronomy, Onsala Space observatory at Chalmers University of technology} (OSO) in February 2007.  The Caltech Submillimeter Observatory (CSO) was used  on March 26, 1994 to detect $J$\,$=$\,8\,$\rightarrow$\,7 line emission. The CSO data are particularly important, not only because it probes warmer and more dense material closer to the central star, but also since it will provide an estimate on the missing short spacing flux in the SMA interferometer data as further discussed in Sect.~\ref{sec_bright}.  
In addition, publically available $J$\,$=$\,3\,$\rightarrow$\,2 and $J$\,$=$\,4\,$\rightarrow$\,3 observations using the JCMT telescope\footnote{Based on observations obtained with the James Clerk Maxwell Telescope, which is operated by the Joint Astronomy Centre in Hilo, Hawaii on behalf of the parent organizations PPARC in the United Kingdom, the National Research Council of Canada and The Netherlands Organization for Scientific Research}, have been obtained.
The observed spectra are presented in Sect.~\ref{model_single}, where direct comparison with model predictions are made, and velocity-integrated intensities are reported in Table~\ref{intensities}. 
The data was reduced in a standard way, by removing a low order baseline and then binned in order to improve the 
signal-to-noise ratio.
The intensity scales are given in main-beam brightness temperature scale ($T_{\mathrm{mb}}$).

The OSO and JCMT observations were made in a dual beamswitch mode, 
where the source is alternately placed in the signal and the reference
beam, using a beam throw of about $11\arcmin$ (OSO) or $2\arcmin$ (JCMT). 
This method produces very flat baselines. The CSO observations were performed using position
switching where off-source data were taken  $\pm$\,180$\arcsec$ away in azimuth.
The raw spectra are stored in the $T_{\mathrm A}^{\star}$  scale and converted to main-beam brightness temperature using $T_{\mathrm{mb}}$\,=\,$T_{\mathrm
A}^{*}/\eta_{\mathrm{mb}}$. $T_{\mathrm A}^{\star}$ is the
antenna temperature corrected for atmospheric attenuation using the
chopper-wheel method, and $\eta_{\mathrm{mb}}$ is the main-beam
efficiency. 
 The adopted beam efficiencies, together with the FWHM of the main beam ($\theta_{\mathrm{mb}}$), for all telescopes and frequencies are given in Table~\ref{intensities}.
The uncertainty in the absolute intensity scale is estimated to be about $\pm 20$\%. In Table~\ref{intensities} also the energy of the upper level involved in the particular transition ($E_u$) is shown, ranging from 4\,K for the $J$\,$=$\,1 level up to 149\,K for the  $J$\,$=$\,8 level, illustrating the potential of these multi-transition observations to probe a large radial range of the CSE (see Sect.~\ref{model_single}).

\section{The H$^{13}$CN brightness distribution}
\label{sec_bright}
The velocity channel maps of the H$^{13}$CN $J$\,$=$\,8\,$\rightarrow$\,7  line emission obtained by the SMA, presented in Fig.~\ref{sma_fig}, show that the brightness distribution towards IRC+10216 appears to have an overall circular symmetry. At this level of sensitivity and resolution no signs of deviations from a homogeneous wind such as mass loss modulations or clumps are evident. There is an offset, on average  ($0\farcs5$, $-0\farcs1$), of the H$^{13}$CN brightness distribution to the adopted source position. The slight offset in the H$^{13}$CN map to the nominal source position likely results from a combination of residual baseline and passband uncertainties.
 
 As interferometers lack sensitivity to large scale emission  it is of interest to ascertain the missing flux of the SMA data. The CSO beam of  10$\arcsec$ recovers all the H$^{13}$CN $J$\,$=$\,8\,$\rightarrow$\,7  line emission from the envelope. Converting the intensity scale in the CSO spectrum  from main-beam-brightness temperature to Jy\,beam$^{-1}$, using a conversion factor of 39\,Jy\,K$^{-1}$, about 840\,Jy is recovered by the CSO beam near the systemic velocity (at $-$\,26\,km\,s$^{-1}$ relative to the LSR). In comparison, the shortest baselines in the interferometer only pick up about 25\% of the total emission.

  \begin{figure*}
      \centering{   
   \includegraphics[width=16cm]{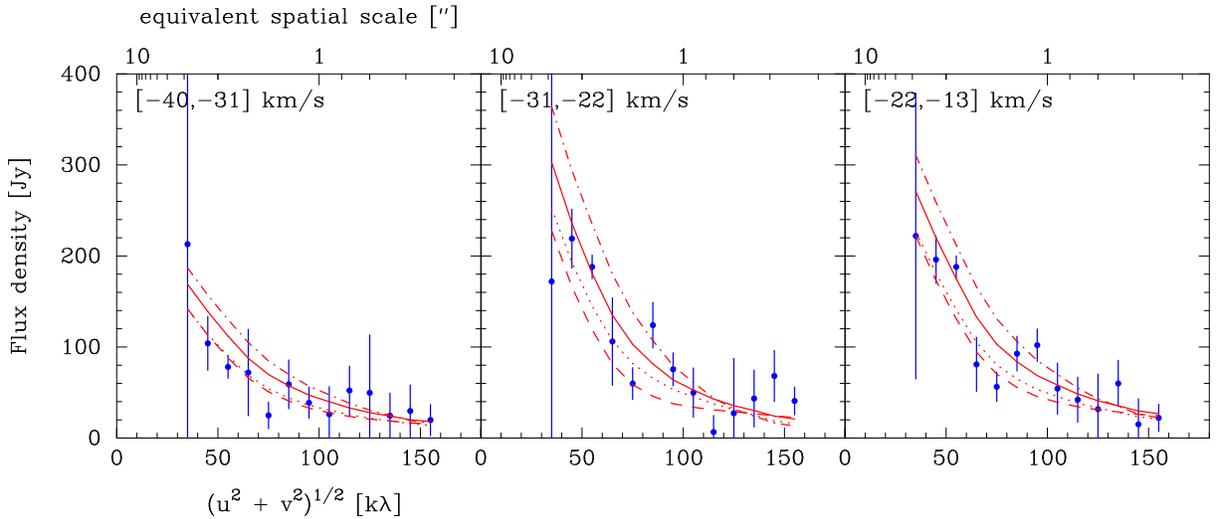}
   \caption{Visibilities, averaged over velocity 
   obtained using SMA interferometer as a function of distance to the phase center in k$\lambda$. 
   (also shown on the upper abscissa is the equivalent spatial resolution in arcseconds, see Sect.~\ref{obs_sma} for its definition and meaning).
   The observations are overlayed by model results that have been sampled in the ($u,v$)-plane in the same way as the observations. The best-fit single-dish model (Model 1) using  a H$^{13}$CN fractional abundance $f_0$\,$=$\,6\,$\times$\,10$^{-7}$  and envelope size $r_{\mathrm{e}}$\,$=$\,3\,$\times$\,10$^{16}$\,cm is indicated by the solid line. Also shown are models using a 50\% lower (Model 2; dashed line) and 100\% higher (Model 3; dash-dotted line) H$_2$ density (corresponding to $\dot{M}$\,$=$\,7.5\,$\times$10$^{-6}$\,M$_{\odot}$\,yr$^{-1}$ and $\dot{M}$\,$=$\,3.0\,$\times$10$^{-5}$\,M$_{\odot}$\,yr$^{-1}$, respectively) within a radial distance of 1.0\,$\times$10$^{15}$\,cm. Also shown is Model 1 including a temperature decrease of 30\% for $r$\,$<$\,3\,$\times$\,10$^{15}$\,cm (Model 4; dotted line). All models are consistent with available single-dish data within 1$\sigma$.}
    \label{sma_uvall}}
      \end{figure*}

The CLEANed image of an extended source with missing short-spacing
data results in an extended depression of negative surface brightness
on which the source emission resides.  This artifact is most evident
near the central velocity channels in Fig.~\ref{sma_fig}, where the real envelope
emission is expected to be the most extended. The negative
feature will further reduce the size and flux of the emission in those channels.
Great care needs to be taken when analysing the data and comparing to model predictions.  For the present set of data we find that the most reliable way is to perform the analysis directly in the $uv$-plane which will also maximize the resolution and sensitivity.

In Fig.~\ref{sma_uvall} the azimuthally-averaged visibility amplitudes are plotted as a function of distance to the phase centre in k$\lambda$.  The emission has been averaged over three velocity bins, each with a width of 9\,km\,s$^{-1}$. There is a clear trend that the visibilities decrease with increased baseline indicating that the emission is well resolved.  There is also a significant spread in the visibilities, in particular around 50\,$-$100\,k$\lambda$. 
The innermost region of IRC+10216 is known to be inhomogeneous to some degree \citep[e.g.,][]{Menshchikov01, Leao06}. While it is tempting to assign the deviation of the observed visibilities from a more smooth decline with baseline separation to departures from a spherically symmetric and homogeneous wind, the quality of the current data set prevents such an interpretation. The implications of the spread in visibilities on the modelling, i.e. uncertainties in the derived physical parameters, are discussed in Sect.~5.2.
 
\section{Excitation analysis}
\label{sect_model}
\subsection{Radiative transfer model}
In order to determine the molecular excitation in the CSE around IRC+10216 a detailed non-LTE radiative transfer code, based on the Monte Carlo method,
was used.  The code is described in detail in \citet{Schoeier01} and has been
benchmarked, to high accuracy, against a wide variety of molecular-line radiative
transfer codes in \citet{Zadelhoff02}.  
The CSE around \object{IRC+10216} is assumed to be spherically symmetric, produced by a constant mass-loss rate ($\dot{M}$), and to expand at a constant velocity ($v_{\mathrm e}$).  In Sect. 5.2 we will relax the assumption of a constant mass-loss rate and allow it to vary in time.

%
%

The physical properties, such as the density, temperature, and kinematic structure prevailing in the circumstellar envelope around \object{IRC+10216} have been determined in \citet{Schoeier00,Schoeier01}, \citet{Schoeier02b}, and \citet{Schoeier06a}, based on radiative transfer modelling of multi-transition CO line observations, from millimetre to IR wavelengths (the reader is referred to these works for more specific details of the input parameters). This model, where the circumstellar envelope is formed by a mass-loss rate of 1.5\,$\times$\,10$^{-5}$\,M$_{\odot}$\,yr$^{-1}$ and expanding at a velocity of 14.0\,km\,s$^{-1}$, is used as input to the H$^{13}$CN excitation analysis.  The inner radius of the model is taken to be $r_0$\,=\,1.7\,$\times$\,10$^{14}$\,cm ($=$\,11\,AU\,$=$\,0\farcs095). The kinetic temperature structure is shown in Sect.~\ref{model_single} where the molecular excitation is discussed.


\subsection{The H$^{13}CN$ abundance distribution}
\label{sma}
The abundance distribution of H$^{13}$CN is assumed to be described by a Gaussian
\begin{equation}
\label{eq_distr}
f(r) = f_0\, \exp \left(-\left(\frac{r}{r_{\mathrm e}}\right)^2 \right),
\end{equation}
where $f$\,$=$\,$n\mathrm{(H^{13}CN)}/n\mathrm{(H_2)}$ is the fractional abundance, i.e., the ratio of the number density of H$^{13}$CN molecules to that of H$_2$ molecules. Here, $f_0$ denotes the photospheric fractional abundance of SiO and it is assumed that some process effectively destroys the molecules at $r$\,$>$\,$\,r_{\mathrm e}$ such as, e.g.,  photodissociation by the ambient interstellar uv-field.  

The excitation analysis includes radiative excitation through the stretching mode at 3\,$\mu$m and in the bending mode at 14\,$\mu$m. The stretching mode at 5\,$\mu$m includes transitions that  are about 300 times weaker and is therefore not included in the analysis. In each of the vibrational levels we include rotational levels up to $J$\,=\,29. Hyperfine splitting of the rotational levels were included only in the $J$\,=\,1 levels (where the splitting is larger than the local turbulent width) and the resulting line overlaps were accurately treated as described in Lindqvist et al.\ (2000). Also, $l$-type doubling in the 14\,$\mu$m transitions was included. Relevant molecular data are summarized in \citet{Schoeier05a} and are made publicly available through the {\em Leiden Atomic and Molecular Database} (LAMDA){\footnote{\tt http://www.strw.leidenuniv.nl/$\sim$moldata}}. 

For \object{IRC+10216}  thermal dust emission provides the main source of infrared photons which excite the vibrational states. The addition of a dust component in the Monte Carlo scheme is described  in \citet{Schoeier02b}. The dust-temperature structure and dust-density profile for \object{IRC+10216} are obtained from 
radiative transfer modelling of the spectral energy distribution using {\em Dusty} \citep{Ivezic97}. Details on the dust modelling can be found in  \citet{Schoeier06a}. The best-fit dust model is obtained for an optical depth at 10\,$\mu$m of 0.9, a dust-condensation temperature of 1200\,K (corresponding to a radial distance of $r_0$\,=\,1.7\,$\times$\,10$^{14}$\,cm and also taken as the inner radius in the H$^{13}$CN excitation analysis) and an effective stellar temperature of 2000\,K. The luminosity is 9600\,L$_{\odot}$, obtained from a period-luminosity relation \citep{Groenewegen96}, and the corresponding distance is 120\,pc. 

   \begin{figure}
   \centering{   
   \includegraphics[width=7cm,angle=-90]{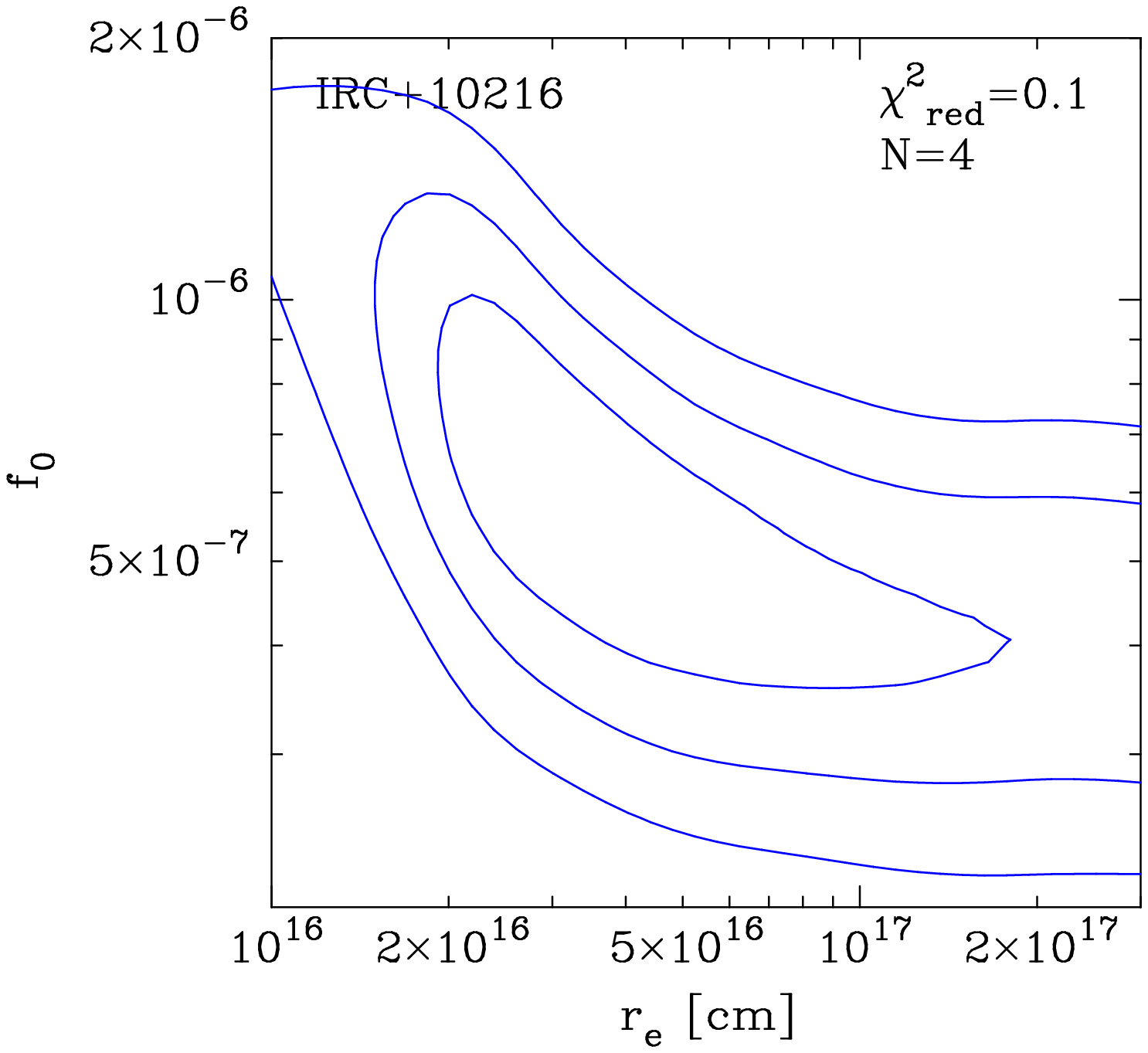}
   \caption{$\chi^2$-map showing the quality of the fit to available single-dish data when varying the adjustable parameters, $f_0$ and $r_{\mathrm{e}}$, in the model. Contours are drawn at the 1, 2, and 3\,$\sigma$ levels. Indicated in the upper right corner are the reduced $\chi^2$ of the best-fit model and the number of observational constrains used, $N$.}
   \label{chi2}}
   \end{figure}

\section{Results and discussion}
\label{sect_discussion}
\subsection{The extended envelope}
\label{model_single}
In order to determine accurate abundances it is important to know the spatial extent of the molecules.  Models where the size of the emitting region ($r_{\mathrm e}$) and the photospheric abundance ($f_0$) in Eq.~\ref{eq_distr} are varied simultaneously have been performed. 
The best fit model is found by minimizing the total $\chi^2$ defined as
\begin{equation}
\label{chi2_sum}
\chi^2_{\mathrm{tot}} = \sum^N_{i=1} \left [ \frac{(I_{\mathrm{mod}}-I_{\mathrm{obs}})}{\sigma}\right ]^2, 
\end{equation} 
where $I$ is the integrated line intensity and $\sigma$ the uncertainty in the measured 
value (usually dominated by the calibration uncertainty of $\pm$20\%), and the summation is done over
$N$ independent observations (here $N$\,$=$\,4). 

First the available observed multi-transition single-dish data (see Fig.~\ref{spectra}) are modelled. These various rotational transitions require different excitation conditions  (see Table~\ref{intensities}) and will probe the CSE at different spatial scales. The energy of the upper level involved in the particular transition ($E_u$) ranges from 4\,K for the $J$\,$=$\,1 level up to 149\,K for the  $J$\,$=$\,8 level. This means that the lines will not behave in the same way to changes of the adjustable parameters $r_{\mathrm{e}}$ and $f_0$ (see Eq.~1) allowing for the possibility to constrain them.
The sensitivity of the line emission to variations of 
$r_{\mathrm{e}}$ and $f_0$ is illustrated in Fig.~\ref{chi2}. We find that it is possible to simultaneously determine both the spatial extent of the H$^{13}$CN molecular envelope, $r_{\mathrm{e}}$\,$=$\,1.0\,$\pm$\,0.8\,$\times$\,10$^{17}$\,cm, and the fractional abundance of  H$^{13}$CN molecules, $f_0$\,$=$\,7.0\,$\pm$\,2.5\,$\times$\,10$^{-7}$.
However,  it is hard to put good constraints on the exact H$^{13}$CN envelope size from modelling of single-dish data alone. 

\begin{figure}
   \centering{
   \includegraphics[width=7cm,angle=-90]{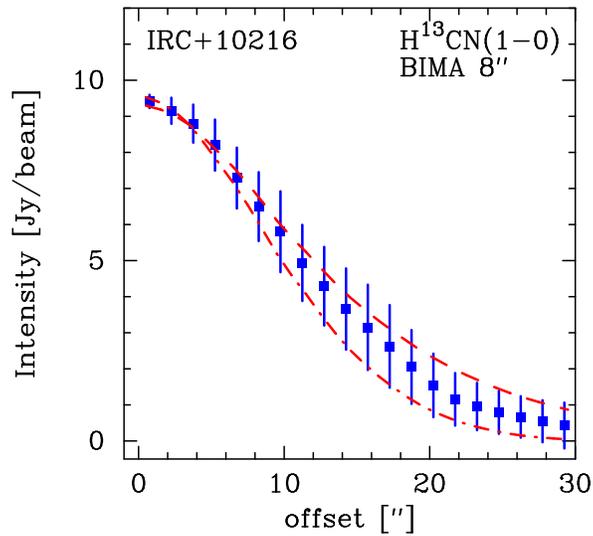}
   \caption{Azimuthally averaged (near the systemic velocity) H$^{13}$CN $J$\,=\,1\,$\rightarrow$\,0 interferometric BIMA observations (Dayal \& Bieging 1995) toward IRC+10216. Models using $f_0$(H$^{13}$CN)\,=\,9$\times$10$^{-7}$ and $r_{\mathrm{e}}$\,$=$\,2.5$\times$10$^{16}$\,cm(dash-dotted line) and  $f_0$(H$^{13}$CN)\,=\,7$\times$10$^{-7}$ and $r_{\mathrm{e}}$\,$=$\,6.0$\times$10$^{16}$\,cm (dashed line) are also shown. The FWHM (8$\arcsec$) of the clean beam is indicated.}
      \label{bima}}
   \end{figure}

 Interferometric observations can help to further constrain the HCN envelope size. In Fig.~\ref{bima} observations of H$^{13}$CN $J$\,=\,1\,$\rightarrow$\,0 line emission towards  IRC+10216 using the BIMA interferometer by \citet{Dayal95} are presented. The spatial resolution of 8$\arcsec$ is enough to constrain the envelope size to within 2.5\,$-$\,6.0\,$\times$\,10$^{16}$\,cm. \citet{Dayal95} in their excitation analysis found a fractional abundance $f_0$(H$^{13}$CN)\,=\,7.8\,$\times$\,10$^{-7}$ and an envelope size $r_{\mathrm{e}}$\,$=$\,2.4\,$\times$\,10$^{16}$\,cm, in good agreement with the present analysis after correcting for their slightly smaller distance and larger mass loss rate.
Lindqvist et al.\ (2000), in their modelling of PdB interferometric HCN $J$\,=\,1\,$\rightarrow$\,0 observations of  IRC+10216, found an envelope size of 4\,$\times\,$10$^{16}$\,cm and $f_0$(HCN)\,=\,5\,$\times$\,10$^{-5}$ which translates to a fractional  H$^{13}$CN abundance of 1\,$\times$\,10$^{-6}$ using a $^{12}$C/$^{13}$C-ratio of 50 (Sch\"oier \& Olofsson 2000) in agreement with the present analysis. It should be noted that Lindqvist et al.\ did  not properly take into account a dust component in their excitation analysis (explaining their somewhat larger fractional abundance), instead they introduced a central, cool (510\,K), blackbody to simulate the dust excitation (see the problems with this simplified strategy discussed by Lindqvist et al.\ 2000). The collisional rate coefficients used were also somewhat different compared to the present analysis.
Additional abundance estimates for IRC+10216 by \citet{Cernicharo96} based on ISO/LWS data and \citet{Cernicharo99} in their analysis of ISO/SWS observations found fractional abundances of HCN in the range 1\,$-$\,3\,$\times$\,10$^{-5}$,  again in good agreement with the present analysis.

Model spectra from the best-fit single-dish model,  i.e. the model with the lowest $\chi^2$ value in Fig.~\ref{chi2}(hereafter Model 1), are directly compared to observations in Fig.~\ref{spectra}. 
The quality of the best-fit model, where $f_0$\,=\,6$\times$10$^{-7}$ and $r_{\mathrm{e}}$\,=\,3$\times$10$^{16}$\,cm, is exceptionally good with a reduced $\chi^2$ ($=$\,$\chi^2_{\mathrm{tot}}$$/$($N-2$)) of 0.1. Also the extra line broadening in the $J$\,=\,1$\rightarrow$0 transition, due to the hyperfine splitting, is very well reproduced. Moreover, the increased amount of self-absorption  due to the higher optical depths with increasing rotational transitions, producing more triangular shaped line profiles and moving the peak of the line towards the red part of the spectrum,  is well reproduced by the model. 

\begin{figure}
      \centering{   
   \includegraphics[width=7cm]{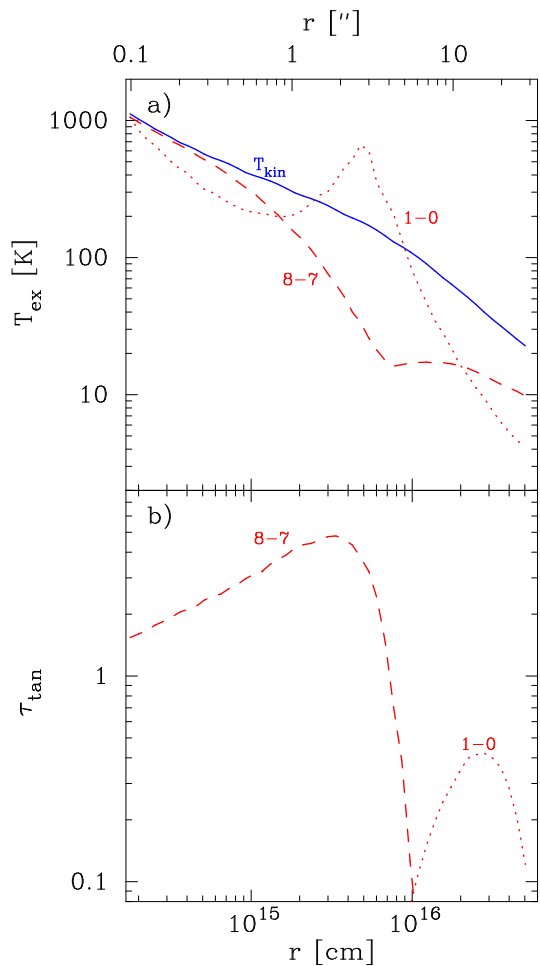}
   \caption{Results for the best-fit single-dish model (Model 1) from the excitation analysis.  a) The kinetic temperature of the gas particles as derived from the CO modelling is shown by the solid line. The dotted and dashed lines gives the excitation temperature of the H$^{13}$CN $J$\,=\,1$\rightarrow$0, $F$\,=\,2$\rightarrow$1 and $J$\,=\,8$\rightarrow$7 transitions, respectively.  b) The dotted and dashed lines gives the tangential optical depth ($\tau_{\mathrm{tan}}$) at the line center of the H$^{13}$CN $J$\,=\,1$\rightarrow$0, $F$\,=\,2$\rightarrow$1 and $J$\,=\,8$\rightarrow$7 transitions, respectively.}
   \label{excitation_fig}}
\end{figure}

Results from the excitation analysis are shown in Fig.~\ref{excitation_fig} for Model~1 where the excitation temperature, $T_{\mathrm{ex}}$, and the tangential optical depth, $\tau_{\mathrm{tan}}$ are plotted as functions of radial distance from the star for the  $J$\,$=$\,1\,$\rightarrow$\,0 and $J$\,$=$\,8\,$\rightarrow$\,7 transitions. It is clear that both lines are formed in non-LTE  conditions as $T_{\mathrm{ex}}$ is far from the kinetic temperature of the gas ($T_{\mathrm{kin}}$, also shown in Fig.~\ref{excitation_fig}) where  $\tau_{\mathrm{tan}}$ is peaking. The  $J$\,$=$\,1\,$\rightarrow$\,0 line emission (shown here for the inherently strongest of the three hyperfine components the $F$\,$=$\,2\,$\rightarrow$\,1 transition) is optically thin where as the $J$\,$=$\,8\,$\rightarrow$\,7 emission is moderately thick. From the excitation analysis it is also clear that the $J$\,$=$\,1\,$\rightarrow$\,0 line emission is mainly produced in the external cooler layers of the H$^{13}$CN envelope and it is thus more sensitive to changes in $r_{\mathrm{e}}$ than the $J$\,$=$\,8\,$\rightarrow$\,7 transition which is more sensitive to $f_0$. As an example, reducing the envelope size in Model~1 by 50\% the $J$\,$=$\,1\,$\rightarrow$\,0 line emission, as observed by the OSO beam, gets 49\% weaker whereas the $J$\,$=$\,8\,$\rightarrow$\,7 line emission, in the CSO beam, is lowered by only 4\%. If instead the fractional abundance in Model~1 is lowered by 50\%, line emission from the $J$\,$=$\,1\,$\rightarrow$\,0 and $J$\,$=$\,8\,$\rightarrow$\,7 transitions are reduced by 34\% and 31\%, respectively. In all, this illustrates the usefulness of multi-transition observations in lifting the degeneracy between envelope size and fractional abundance which otherwise severely hampers an excitation analysis when no direct information of the spatial distribution of the molecules exist.

In the excitation analysis it is important to include also vibrationally excited states. Retaining only the ground vibrational state in Model~1 lowers the line intensities of the observed transitions by 42\% on average. This means that in a multi-transitional analysis, since the various transitions behave in a different manner, not only the derived fractional abundance will be severely affected but also, to a lesser extent, the physical size of the H$^{13}$CN envelope. 

\subsection{The inner wind}
\label{inner_wind}
The interferometric SMA observations of H$^{13}$CN $J$\,=\,8$\rightarrow$7 line emission provide additional constraints on the H$^{13}$CN modelling. From the excitation properties of the $J$\,=\,8$\rightarrow$7 transition discussed in Sect.~\ref{model_single} it is clear that this line will mainly provide information on the warmer more dense circumstellar material closer to the star, typically within $5\arcsec$ in radius. This is further accentuated by the lack of short spacing information, and hence loss of sensitivity to extended emission, in the set of data analysed here (see Sect.~\ref{obs_sma}). 

 The visibilities as a function of baseline shown in Fig.~\ref{sma_uvall} should be regarded as the best constraint on the model as it is then tested over a large range of spatial scales. The inversion process from the ($u,v$)-plane to the image domain will also suffer from artefacts introduced by the large amount of missing flux, and although this should be the case also when analysing the model results, this may enhance deviations from the model.

Based on the best-fit model from the single-dish analysis a brightness map of the emission around IRC+10216 was produced. To this map the same spatial filtering as present in the observations was applied and azimuthal averages produced in the ($u,v$)-plane. The observed visibilities can then be directly compared with those from the model.  As shown in Fig.~\ref{sma_uvall}  the best-fit single-dish model (Model 1; dotted line) provides a reasonable fit to the observed visibilities over the full range of baseline separations, both for velocities near the center and close to the edges of the line. This also means that the data is consistent with a constant expansion at the terminal velocity (14\,km\,s$^{-1}$) of the wind. The spread in visibilities that is observed can be accounted for by modulating the mass-loss rate. In Fig.~\ref{sma_uvall}  models with a 50\% lower (Model 2; dashed line) and 100\% higher (Model 3; dash-dotted line) H$_2$ density (corresponding to $\dot{M}$\,$=$\,7.5\,$\times$10$^{-6}$\,M$_{\odot}$\,yr$^{-1}$ and $\dot{M}$\,$=$\,3.0\,$\times$10$^{-5}$\,M$_{\odot}$\,yr$^{-1}$, respectively), within a radial distance of 1.0\,$\times$\,10$^{15}$\,cm are shown. The fits to the single-dish data are still within the 1$\sigma$ limit. With the uncertainties of the observations in mind, we find the single-dish data and the SMA interferometric data to be consistent with an H$^{13}$CN abundance distribution described by Eq.~\ref{eq_distr} with $f_0$\,$=$\,4\,$\times$\,10$^{-7}$ and $r_{\mathrm{e}}$\,$=$\,4\,$\times$\,10$^{16}$\,cm.
 
The SMA data can not provide an estimate of the H$^{13}$CN envelope size, within reasonable limits, due to the notorious problem of filtering out of large scale weak emission that hampers interferometers and the fact that the $J$\,=\,8\,$\rightarrow$\,7 emission is mostly excitation limited rather than photodissociation limited (see Sect.~5.1).

  \begin{figure}
    \label{cooling_fig}
      \centering{   
   \includegraphics[width=7cm,angle=-90]{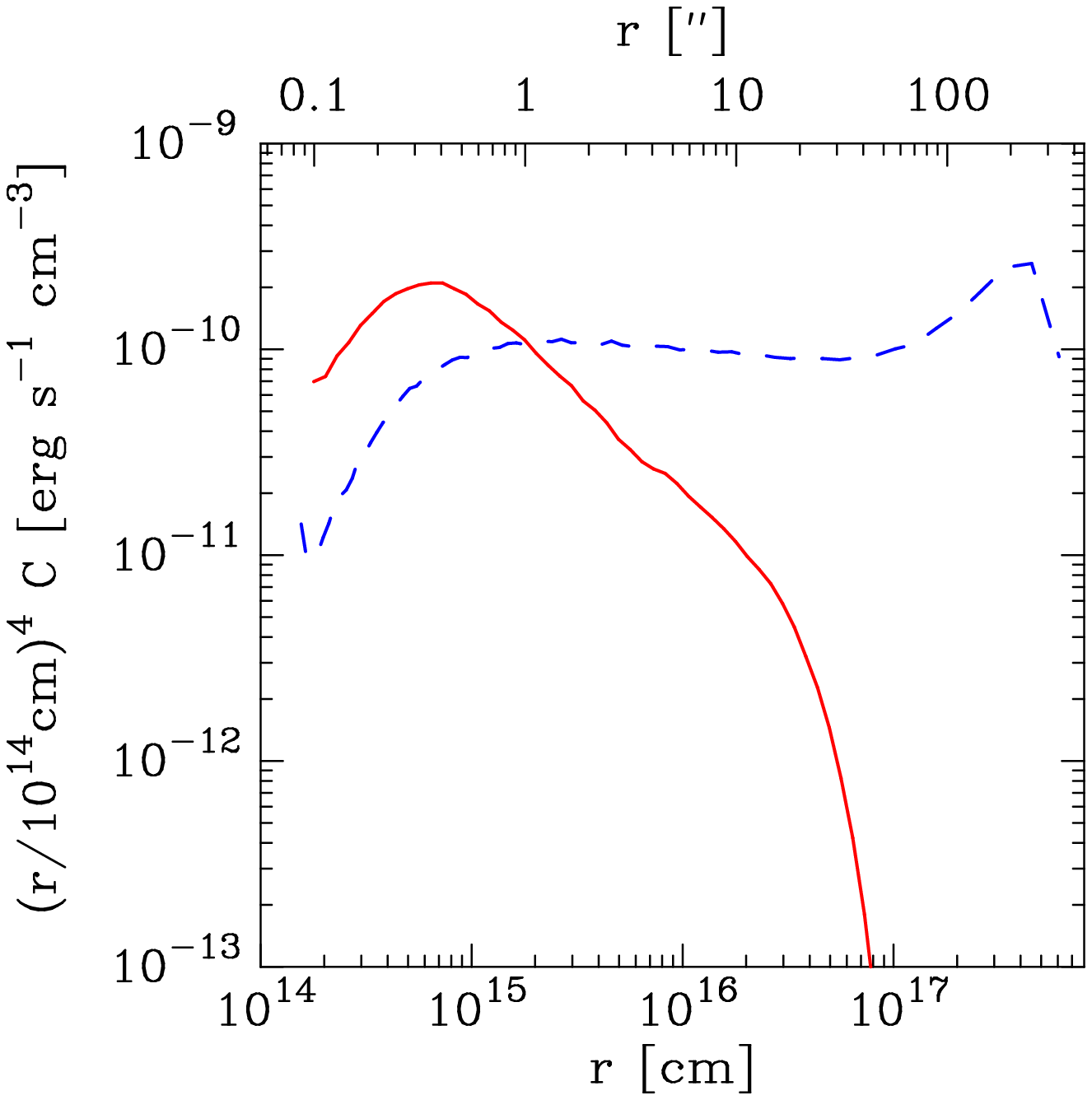}
   \caption{HCN line cooling obtained from the best-fit single-dish model (Model 1; solid line) compared with that obtained from CO (dashed line). The same underlying physical structure has been used in the two cases.}}
      \end{figure}

\subsection{HCN line cooling}
It has been suggested that HCN line emission could be an important coolant in the inner parts of circumstellar envelopes around carbon stars \citep{Groenewegen94,Cernicharo96}. 
In Fig.~\ref{cooling_fig}  we show the estimated total HCN line cooling from our best-fit single dish model (Model 1; solid line). The line cooling from H$^{13}$CN can be self-consistently calculated once the excitation is known from 
\begin{equation} 
C = n_{\mathrm{H}_2} \sum_l\sum_{u>l} (n_u\gamma_{ul}-n_l\gamma_{lu})h\nu_{ul},
\end{equation}
where $n_i$ denotes the level populations of H$^{13}$CN, $\gamma_{ul}$ the collisional rate coefficients, $n_{\mathrm{H}_2}$ is the H$_2$ number density, and $C$ is the emissivity per molecule in erg\,s$^{-1}$\,cm$^{-3}$.
In order to obtain the total HCN cooling this result was multiplied by the isotopic ratio $^{12}$C/$^{13}$C\,$=$\,50. Within a radial distance of about 2\,$\times$\,10$^{15}$\,cm ($\approx$\,1\arcsec) HCN line emission dominates over that from CO. Thus a lower kinetic temperature is expected. To simulate this effect the kinetic temperature structure was lowered by 30\% in this region.  This has no effect for the single-dish data given their larger beam-sizes ($\gtrsim$\,10\arcsec) but lowers the intensity of  the H$^{13}$CN $J$\,=\,8\,$\rightarrow$\,7 line emission from the region where HCN line cooling is effective, still providing a reasonable fit to the observations (Fig.~\ref{sma_uvall}; dotted line; Model 4).

\subsection{Comparison with chemical models}
 Recent studies of SiO line emission around IRC+10216 by \citet{Schoeier06c} and \citet{Schoeier06a} have shown that non-equilibrium chemical processes are required in order to explain the very high amounts (almost two orders of magnitudes higher than expected from an equilibrium chemistry) of SiO present in or close to the photosphere. In addition, at $\sim$\,5 stellar radii there is a drastic drop in the fractional abundance by about an order of magnitude explained by effective adsorption onto dust grains. The effect of shocks is strongly dependent on the C/O-ratio and in a carbon star such as IRC+10216 the minor species, such as SiO, are expected to be much more strongly affected than the most abundant ones, such as HCN. Also, SiO is very refractory compared to HCN, which means that SiO will condense at higher temperatures much closer to the star, where the densities are significantly higher, with the result that the fractional abundance of HCN in the gas phase is expected to be much less affected by this process.

LTE stellar atmosphere models predict  that the HCN fractional abundance in carbon stars is in the range $\approx$\,1\,$-$\,5\,$\times$\,10$^{-5}$ \citep[e.g.,][]{Willacy98, Cherchneff06} which agrees well with the value of approximately 2\,$\times$\,10$^{-5}$ derived for IRC+10216 in the present analysis and with circumstellar values for carbon stars in general \citep{Lindqvist00, Woods03}. This would suggest that HCN is formed in the photosphere under near LTE conditions.

Recent non-equilibrium chemistry models of the dynamically pulsating atmosphere of AGB stars indicate that the passage of shock waves can have a significant effect on the abundance of HCN close to the photosphere. It is likely that both the C/O-ratio and the shock parameters can have a large effect on the processing of HCN in the innermost part of the wind (1\,$-$\,5 stellar radii). The fractional abundance of HCN just before dust formation sets in is estimated to be $\approx$\,4\,$-$\,6\,$\times$\,10$^{-6}$ for carbon stars \citep{Willacy98, Cherchneff06}, i.e., up to an order of magnitude lower than those predicted from equilibrium models. For an extreme carbon star such as IRC+10216, where the C/O-ratio is $\approx$\,1.5, the initially high HCN fractional abundance (relative to H$_2$) gets reduced to about  4.4\,$\times$\,10$^{-6}$ at 5 stellar radii (before dust condensation), i.e., a factor of about five lower than the circumstellar value obtained in the present analysis. However, given the uncertainties involved in both the chemical models and the derived HCN fractional abundance from observations we find that it  is not possible to rule out the importance of non-equilibrium chemical processes in the formation of HCN in the (extended) atmospheres of IRC+10216. Possibly, a statistical study based on a large survey of circumstellar HCN line emission, from AGB stars with varying photospheric (e.g., C/O-ratio) and circumstellar (e.g., mass-loss rate) properties, could provide better constraints on the chemical models. Such work is currently underway (Sch\"oier et al., in prep.). The apparent consistency between LTE stellar atmosphere models and the derived circumstellar HCN fractional abundance in IRC+10216, together with the suggestion that non-equilibrium chemical processes in the extended atmosphere will serve to reduce the photospheric abundance, indicates that few HCN molecules are incorporated into dust grains.

\section{Conclusions}
A detailed excitation analysis, based around a non-LTE and non-local radiative transfer code, has been performed for H$^{13}$CN in the circumstellar envelope around the extreme carbon star IRC+10216. New sub-arcsecond interferometric observations of H$^{13}$CN $J$\,=\,8\,$\rightarrow$\,7 line emission, performed using the SMA, put constraints on the HCN abundance distribution on scales $\lesssim$\,$5$\,$\times$\,10$^{15}$\,cm  ($=$\,330\,AU\,$=$\,2\farcs8). In combination with multi-transition single-dish observations, probing the emission on larger spatial scales, it is found that the H$^{13}$CN fractional abundance remains more or less unaffected in the envelope at about 4\,$\times$\,10$^{-7}$ until photodissociation sets in limiting the size to about 4\,$\times$\,10$^{16}$\,cm  ($=$\,2700\,AU\,$=$\,22$\arcsec$). In the excitation analysis of HCN it is of utmost importance that radiative excitation through the vibrationally excited states is properly accounted for. Also, HCN provides effective line cooling in the inner part of the envelope ($\lesssim$\,2\,$\times$\,10$^{15}$\,cm) that is higher than that from CO. The fractional abundance of HCN derived in the circumstellar envelope of IRC+10216 is consistent with model atmosphere predictions  and suggests that non-equilibrium chemical processes in the extended pulsating atmosphere has no drastic net effect in regulating the amount of HCN molecules that enters the outer wind.
 
\begin{acknowledgements}
The authors are very grateful to Mark Gurwell for assistance during the SMA observing run. We thank an anonymous referee for constructive comments that helped improve the paper. Hans Olofsson is thanked for useful comments on the manuscript. FLS acknowledges financial support from the Swedish Research Council. JHB acknowledges partial support from NSF grant AST-0307687 to the University of Arizona. 
\end{acknowledgements}
%


{\it Facilities:} \facility{SMA}.


\begin{thebibliography}{41}
\expandafter\ifx\csname natexlab\endcsname\relax\def\natexlab#1{#1}\fi

\bibitem[{{Ag{\'u}ndez} \& {Cernicharo}(2006)}]{Agundez06}
{Ag{\'u}ndez}, M. \& {Cernicharo}, J. 2006, \apj, 650, 374

\bibitem[{{Bieging} {et~al.}(2000){Bieging}, {Shaked}, \&
  {Gensheimer}}]{Bieging00}
{Bieging}, J.~H., {Shaked}, S., \& {Gensheimer}, P.~D. 2000, \apj, 543, 897

\bibitem[{{Bieging} \& {Tafalla}(1993)}]{Bieging93}
{Bieging}, J.~H. \& {Tafalla}, M. 1993, \aj, 105, 576

\bibitem[{{Cernicharo} {et~al.}(1996){Cernicharo}, {Barlow},
  {Gonzalez-Alfonso}, {Cox}, {Clegg}, {Nguyen-Q-Rieu}, {Omont}, {Guelin},
  {Liu}, {Sylvester}, {Lim}, {Griffin}, {Swinyard}, {Unger}, {Ade}, {Baluteau},
  {Caux}, {Cohen}, {Emery}, {Fischer}, {Furniss}, {Glencross}, {Greenhouse},
  {Gry}, {Joubert}, {Lorenzetti}, {Nisini}, {Orfei}, {Pequignot}, {Saraceno},
  {Serra}, {Skinner}, {Smith}, {Towlson}, {Walker}, {Armand}, {Burgdorf},
  {Ewart}, {di Giorgio}, {Molinari}, {Price}, {Sidher}, {Texier}, \&
  {Trams}}]{Cernicharo96}
{Cernicharo}, J., {Barlow}, M.~J., {Gonzalez-Alfonso}, E., {et~al.} 1996, \aap,
  315, L201

\bibitem[{{Cernicharo} {et~al.}(1999){Cernicharo}, {Yamamura},
  {Gonz{\'a}lez-Alfonso}, {de Jong}, {Heras}, {Escribano}, \&
  {Ortigoso}}]{Cernicharo99}
{Cernicharo}, J., {Yamamura}, I., {Gonz{\'a}lez-Alfonso}, E., {et~al.} 1999,
  \apjl, 526, L41

\bibitem[{{Cherchneff}(2006)}]{Cherchneff06}
{Cherchneff}, I. 2006, \aap, 456, 1001

\bibitem[{{Cherchneff} \& {Glassgold}(1993)}]{Cherchneff93b}
{Cherchneff}, I. \& {Glassgold}, A.~E. 1993, \apjl, 419, L41

\bibitem[{{Cherchneff} {et~al.}(1993){Cherchneff}, {Glassgold}, \&
  {Mamon}}]{Cherchneff93a}
{Cherchneff}, I., {Glassgold}, A.~E., \& {Mamon}, G.~A. 1993, \apj, 410, 188

\bibitem[{{Dayal} \& {Bieging}(1993)}]{Dayal93}
{Dayal}, A. \& {Bieging}, J.~H. 1993, \apjl, 407, L37

\bibitem[{{Dayal} \& {Bieging}(1995)}]{Dayal95}
{Dayal}, A. \& {Bieging}, J.~H. 1995, \apj, 439, 996

\bibitem[{{Doty} \& {Leung}(1998)}]{Doty98}
{Doty}, S.~D. \& {Leung}, C.~M. 1998, \apj, 502, 898

\bibitem[{{Fong} {et~al.}(2003){Fong}, {Meixner}, \& {Shah}}]{Fong03}
{Fong}, D., {Meixner}, M., \& {Shah}, R.~Y. 2003, \apjl, 582, L39

\bibitem[{{Fong} {et~al.}(2006){Fong}, {Meixner}, {Sutton}, {Zalucha}, \&
  {Welch}}]{Fong06}
{Fong}, D., {Meixner}, M., {Sutton}, E.~C., {Zalucha}, A., \& {Welch}, W.~J.
  2006, \apj, 652, 1626

\bibitem[{{Ford} {et~al.}(2004){Ford}, {Neufeld}, {Schilke}, \&
  {Melnick}}]{Ford04}
{Ford}, K.~E.~S., {Neufeld}, D.~A., {Schilke}, P., \& {Melnick}, G.~J. 2004,
  \apj, 614, 990

\bibitem[{{Groenewegen}(1994)}]{Groenewegen94}
{Groenewegen}, M.~A.~T. 1994, \aap, 290, 531

\bibitem[{{Groenewegen} \& {Whitelock}(1996)}]{Groenewegen96}
{Groenewegen}, M.~A.~T. \& {Whitelock}, P.~A. 1996, \mnras, 281, 1347

\bibitem[{{H{\" o}gbom}(1974)}]{Hoegbom74}
{H{\" o}gbom}, J.~A. 1974, \aaps, 15, 417

\bibitem[{{Hasegawa} {et~al.}(2006){Hasegawa}, {Kwok}, {Koning}, {Volk},
  {Justtanont}, {Olofsson}, {Sch\"oier}, {Winnberg}, {Nyman}, {Frisk},
  {Hjalmarson}, {Olberg}, \& {Sandqvist}}]{Hasegawa06}
{Hasegawa}, T., {Kwok}, S., {Koning}, N., {et~al.} 2006, \apj, 637, 791

\bibitem[{{Ho} {et~al.}(2004){Ho}, {Moran}, \& {Lo}}]{Ho04}
{Ho}, P.~T.~P., {Moran}, J.~M., \& {Lo}, K.~Y. 2004, \apjl, 616, L1

\bibitem[{{Ivezi{\' c}} \& {Elitzur}(1997)}]{Ivezic97}
{Ivezi{\' c}}, {\v{Z}}. \& {Elitzur}, M. 1997, MNRAS, 287, 799

\bibitem[{{Keady} \& {Ridgway}(1993)}]{Keady93}
{Keady}, J.~J. \& {Ridgway}, S.~T. 1993, \apj, 406, 199

\bibitem[{{Le{\~a}o} {et~al.}(2006){Le{\~a}o}, {de Laverny}, {M{\'e}karnia},
  {de Medeiros}, \& {Vandame}}]{Leao06}
{Le{\~a}o}, I.~C., {de Laverny}, P., {M{\'e}karnia}, D., {de Medeiros}, J.~R.,
  \& {Vandame}, B. 2006, \aap, 455, 187

\bibitem[{{Lindqvist} {et~al.}(2000){Lindqvist}, {Sch{\" o}ier}, {Lucas}, \&
  {Olofsson}}]{Lindqvist00}
{Lindqvist}, M., {Sch{\" o}ier}, F.~L., {Lucas}, R., \& {Olofsson}, H. 2000,
  \aap, 361, 1036

\bibitem[{{Lucas} {et~al.}(1995){Lucas}, {Guelin}, {Kahane}, {Audinos}, \&
  {Cernicharo}}]{Lucas95}
{Lucas}, R., {Guelin}, M., {Kahane}, C., {Audinos}, P., \& {Cernicharo}, J.
  1995, \apss, 224, 293

\bibitem[{{Melnick} {et~al.}(2001){Melnick}, {Neufeld}, {Ford}, {Hollenbach},
  \& {Ashby}}]{Melnick01}
{Melnick}, G.~J., {Neufeld}, D.~A., {Ford}, K.~E.~S., {Hollenbach}, D.~J., \&
  {Ashby}, M.~L.~N. 2001, \nat, 412, 160

\bibitem[{{Men'shchikov} {et~al.}(2001){Men'shchikov}, {Balega}, {Bl{\"o}cker},
  {Osterbart}, \& {Weigelt}}]{Menshchikov01}
{Men'shchikov}, A.~B., {Balega}, Y., {Bl{\"o}cker}, T., {Osterbart}, R., \&
  {Weigelt}, G. 2001, \aap, 368, 497

\bibitem[{{Millar} {et~al.}(2001){Millar}, {Flores}, \& {Markwick}}]{Millar01}
{Millar}, T.~J., {Flores}, J.~R., \& {Markwick}, A.~J. 2001, \mnras, 327, 1173

\bibitem[{{Millar} \& {Herbst}(1994)}]{Millar94}
{Millar}, T.~J. \& {Herbst}, E. 1994, \aap, 288, 561

\bibitem[{{Millar} {et~al.}(2000){Millar}, {Herbst}, \& {Bettens}}]{Millar00}
{Millar}, T.~J., {Herbst}, E., \& {Bettens}, R.~P.~A. 2000, \mnras, 316, 195

\bibitem[{{Sch{\" o}ier} \& {Olofsson}(2000)}]{Schoeier00}
{Sch{\" o}ier}, F.~L. \& {Olofsson}, H. 2000, \aap, 359, 586

\bibitem[{{Sch{\" o}ier} \& {Olofsson}(2001)}]{Schoeier01}
{Sch{\" o}ier}, F.~L. \& {Olofsson}, H. 2001, \aap, 368, 969

\bibitem[{{Sch{\" o}ier} {et~al.}(2002){Sch{\" o}ier}, {Ryde}, \&
  {Olofsson}}]{Schoeier02b}
{Sch{\" o}ier}, F.~L., {Ryde}, N., \& {Olofsson}, H. 2002, \aap, 391, 577

\bibitem[{{Sch{\" o}ier} {et~al.}(2005){Sch{\" o}ier}, {van der Tak}, {van
  Dishoeck}, \& {Black}}]{Schoeier05a}
{Sch{\" o}ier}, F.~L., {van der Tak}, F.~F.~S., {van Dishoeck}, E.~F., \&
  {Black}, J.~H. 2005, \aap, 432, 369

\bibitem[{{Sch{\"o}ier} {et~al.}(2006{\natexlab{a}}){Sch{\"o}ier}, {Fong},
  {Olofsson}, {Zhang}, \& {Patel}}]{Schoeier06c}
{Sch{\"o}ier}, F.~L., {Fong}, D., {Olofsson}, H., {Zhang}, Q., \& {Patel}, N.
  2006{\natexlab{a}}, \apj, 649, 965

\bibitem[{{Sch{\"o}ier} {et~al.}(2006{\natexlab{b}}){Sch{\"o}ier}, {Olofsson},
  \& {Lundgren}}]{Schoeier06a}
{Sch{\"o}ier}, F.~L., {Olofsson}, H., \& {Lundgren}, A.~A. 2006{\natexlab{b}},
  \aap, 454, 247

\bibitem[{{Tenenbaum} {et~al.}(2006){Tenenbaum}, {Apponi}, {Ziurys},
  {Ag{\'u}ndez}, {Cernicharo}, {Pardo}, \& {Gu{\'e}lin}}]{Tenenbaum06}
{Tenenbaum}, E.~D., {Apponi}, A.~J., {Ziurys}, L.~M., {et~al.} 2006, \apjl,
  649, L17

\bibitem[{{van Zadelhoff} {et~al.}(2002){van Zadelhoff}, {Dullemond}, {van der
  Tak}, {Yates}, {Doty}, {Ossenkopf}, {Hogerheijde}, {Juvela}, {Wiesemeyer}, \&
  {Sch{\" o}ier}}]{Zadelhoff02}
{van Zadelhoff}, G.-J., {Dullemond}, C.~P., {van der Tak}, F.~F.~S., {et~al.}
  2002, \aap, 395, 373

\bibitem[{{Willacy}(2004)}]{Willacy04}
{Willacy}, K. 2004, \apjl, 600, L87

\bibitem[{{Willacy} \& {Cherchneff}(1998)}]{Willacy98}
{Willacy}, K. \& {Cherchneff}, I. 1998, \aap, 330, 676

\bibitem[{{Woods} {et~al.}(2003){Woods}, {Sch{\" o}ier}, {Nyman}, \&
  {Olofsson}}]{Woods03}
{Woods}, P.~M., {Sch{\" o}ier}, F.~L., {Nyman}, L.-{\AA}., \& {Olofsson}, H.
  2003, \aap, 402, 617

\bibitem[{{Young} {et~al.}(2004){Young}, {Hunter}, {Wilner}, {Gurwell},
  {Barrett}, {Blundell}, {Christensen}, {Fong}, {Hirano}, {Ho}, {Liu}, {Lo},
  {Martin}, {Matsushita}, {Moran}, {Ohashi}, {Papa}, {Patel}, {Patt}, {Peck},
  {Qi}, {Saito}, {Schinckel}, {Shinnaga}, {Sridharan}, {Takakuwa}, {Tong}, \&
  {Trung}}]{Young04}
{Young}, K.~H., {Hunter}, T.~R., {Wilner}, D.~J., {et~al.} 2004, \apjl, 616,
  L51

\end{thebibliography}
\end{document}